\documentclass[twocolumn,twoside]{article}
\usepackage{graphics}
\usepackage[T1,T2A]{fontenc}  
\usepackage[utf8]{inputenc}
\usepackage[ukrainian,english]{babel}

\newcommand{\hb}{\\ \hspace*{2ex}}

\begin{document}

\title{SIMULATION OF LARGE-SCALE STRUCTURE OF UNIVERSE BY GAUSSIAN RANDOM FIELDS}
\author{V.\,V. Voitsekhovskiy, A.\,V. Tugay \\[2mm] 
 Taras Shevchenko National University of Kyiv,\hb
 Kyiv, Ukraine,  {\em tugay.anatoliy@gmail.com}\\
}
\date{}
\maketitle
%
%
ABSTRACT. Large-scale structure of Universe includes galaxy clusters connected by filaments. Voids occupy the rest of cosmic volume. The search of any dependencities in filament structure can give answer to more general questions about origin of structures in the Universe. 
This becomes possible because, according to current picture of Universe, one could simulate the evolution of Universe until its very beginning or vice versa. 
One of the theories which describe the shape of large-scale structures is adiabatic Zeldovich theory. 
This theory explain three-dimensional galaxy distribution as a set of thin pancakes which were formed from hot primordial gas under own gravitational pressure in the cosmological period of acоustic oscillations. 
According to cosmological hydrodynamical theories a number of computer simulations of LSS were performed to describe its properties.
In this work we consider alternative variant of simulating the distribution of matter that is very similar to real.
We simulated two-dimensional galaxy distribution on the sky using random distributions of clusters and single galaxies.
The main assumption was that matter clusterised to initial density fluctuations with uniform distribution.
According to Zeldovich theory, low-dimensional anisotropies should increase, that corresponds to appearance of filaments in 2D case.
Thus we generated a net of filaments between clusters with certain length limits.
Real galaxy distribution was simulated by random changing galaxy positions in filaments and clusters.
We generated radial distributions of galaxies in clusters taking into account the surrounding and add uniform distribution of isolated galaxies in voids.
Our model has been coordinated with SDSS galaxy distribution with using two-point angular correlation function.
Parameters of random distributions were found for the case of equality of correlation function slope for the model and for observational data. \\[1mm]
МОДЕЛЮВАННЯ ВЕЛИКОМАСШТАБНОЇ СТРУКТУРИ ВСЕСВІТУ ЗА ДОПОМОГОЮ ГАУССІВСЬКИХ ВИПАДКОВИХ ПОЛІВ.
Великомасштабна структура Всесвіту включає скупчення галактик, що з'єднані філаментами. Войди займають решту космічного простору. Пошуки закономірностей у структурі філаментів можуть дати відповіді на питання щодо походження структур у Всесвіті. Це можливо завдяки чисеньним моделюванням Всесвіту від його народження до сучасності і навпаки. Форми великомасштабних структур пояснюються у адіабатичній теорії Зельдовича. В цій теорії тривимірний розподіл галактик утворює сукупність тонких "млинців", що формуються з первинного гарячого газу під дією гравітації у космологічний період домінування акустичних осциляцій. Для опису великомасштабних струкутур були виконані багаточисленні комп'ютерні симуляції з використанням космологічних гідродинамічних теорій. У даній роботі ми пропонуємо альтернативний метод моделювання розподілу матерії, що дає результат досить близький до реального. Був змодельований двовимірний розподіл галактик по небу на основі випадкових розподілій скупчень і окремих галактик. Використано припущення про те, що матерія збирається у скупчення до первинних флуктуацій густини, що мають рівномірний розподіл. Відповідно до теоії Зельдовича, анізотропія структур нижчої розмірності повинна зростати, що відповідає утворенню філаментів у двовимірному випадку. Тому у даній роботі пропонується генерування мережі філаментів між скупченнями з визначеними обмеженнями довжини. Реальний розподіл галактик був змодельований випадковими зміщеннями положень галактик у філаментах і скупченнях.  
Радіальні розподіли галактик у скупченнях були згенеровані з врахуванням оточення;
додатково був згенерований однорідний розподіл ізольованих галактик.
Нова модель була узгоджена з розподілом галактик SDSS шляхом порівняння утової двоточкової кореляційної функції. Параметри випадкових розподілів були визначені для випадку рівності нахилу кореляційної функції моделі і спостережних даних.\\[1mm]
{\bf Ключові слова}: великомасштабна структура Всесвіту, галактики, філаменти.
\\[2mm]
%
%
{\bf 1. Introduction}\\[1mm]

Large-scale structure of Universe represents clusters of galaxies divided by large voids. In general it appears as needles-like structure.
Non-Hubble galaxy velocities lead to distortion of galaxy distribution in redshift space. A handy way to study LSS is to consider 2D concentric layers (Tugay, 2012. Thickness of the layer is 100 Mpc that are responsible for one galaxy supercluster.
In this work we performed two-dimensional simulation of large-scale structure of Universe in one 100 Mpc-thick layer corresponding to single supercluster. 
A combination of random functions was selected and implemented to simulate SDSS-like galaxy distribution. Resulting distribution was tested with two-point angular correlation function.\\[2mm]

{\bf 2. Description of new method of LSS simulation}\\[1mm]
 We started with random uniform distribution of points which are considered as galaxy groups or clusters. The points were connected with straight filaments if the distance between them lies in some definite range. Then we generated spherical distribution of galaxies in each cluster. The number of galaxies in cluster was selected as power function of number of filaments pointing to this cluster. We find that isolated galaxies are not generated by this method so we generated additional random distribution of galaxies in all volume.

Physical argumentation of our method is the following. Starting points simulate the largest initial density fluctuations. They grow fastest and become interconnected by filaments. If there are a region with a number of fluctuations with many connections that means that it is large overdensity and there should be larger number of galaxies (Doroshkevich, 1980; Kim, 2009). Also there should be small density fluctuations that form isolated galaxies. So we can divide galaxies by three groups: galaxies in clusters, galaxies in filaments and isolated galaxies. If we will exclude from real volume observed clusters and isolated galaxies, only filament galaxies will remain.This gives us general picture of structure and interconnection of different elements of large-scale structure.\\[2mm]

\begin{figure}[h] 
\resizebox{1.0\hsize}{!}
{\includegraphics{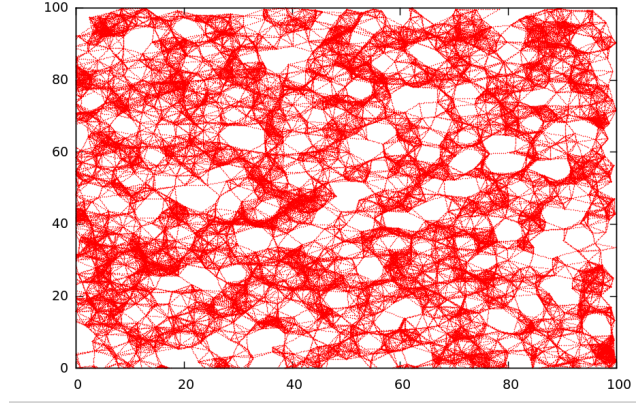}}
\label{fig1}
\caption{Hexagonal point grid or other simple distribution.}
\end{figure}

\begin{figure}[h] 
\resizebox{1.0\hsize}{!}
{\includegraphics{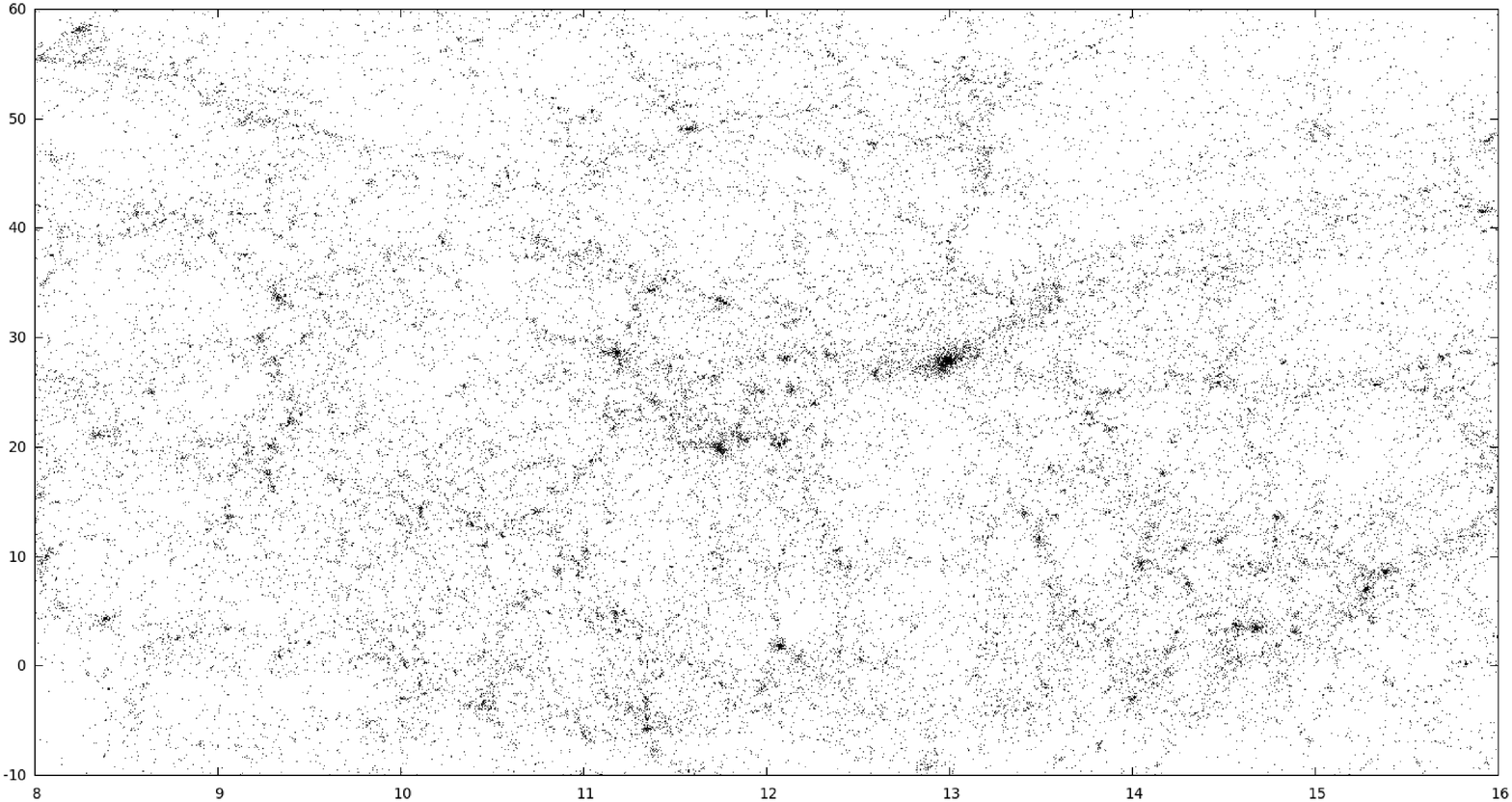}}
\label{fig2}
\caption{Real distribution of galaxies.}
\end{figure}

\begin{figure}[h] 
\resizebox{1.0\hsize}{!}
{\includegraphics{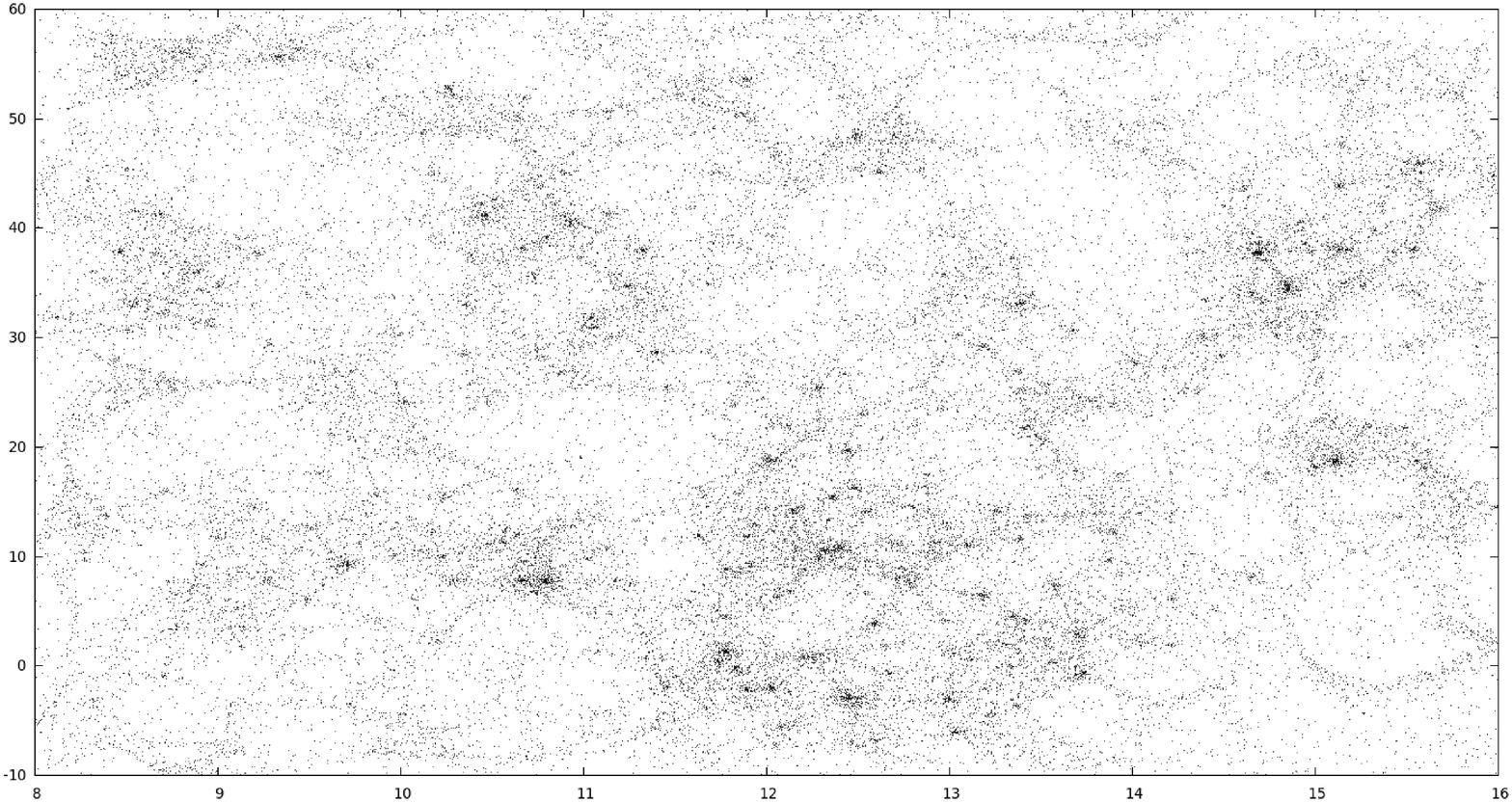}}
\label{fig3}
\caption{Our model of galaxy distribution.}
\end{figure}

\begin{figure}[h] 
\resizebox{1.0\hsize}{!}
{\includegraphics{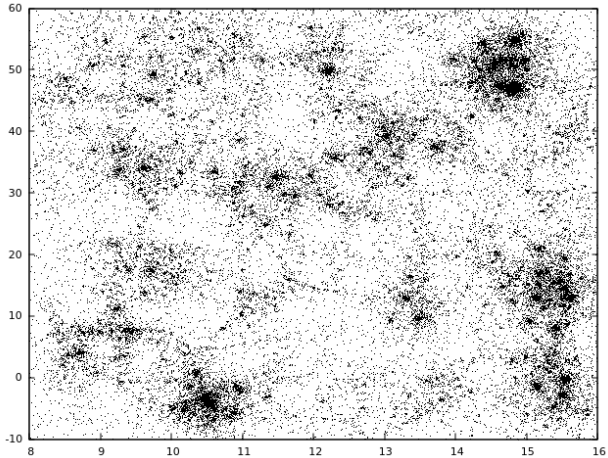}}
\label{fig4}
\caption{Model with reduced coefficients.}
\end{figure}

\begin{figure}[h] 
\resizebox{1.0\hsize}{!}
{\includegraphics{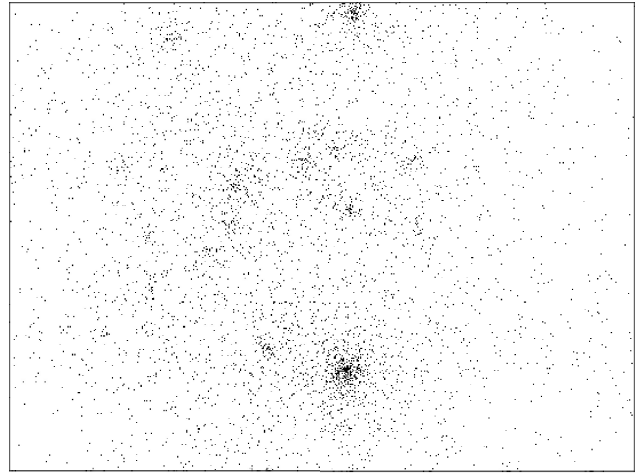}}
\label{fig5}
\caption{Enlarged part of distribution.}
\end{figure}

{\bf 3. Realisation of new method}\\[1mm]

We chose a layer with a thickness of 100 Mpc and filled it with galaxy clusters, filaments and isolated galaxies. Galaxies in clusters and filaments hold a Gaussian distribution. The resulting model has the following parameters. 

1. $r_1$ and $r_2$ - minimal and maximal distances for filament construction. Filament is constructed if distance between cluster is less then $r_2$ and larger then $r_1$. This parameters were selected proportional to average distance $r$ between clusters. Initial bounding distances were selected as following: 
\begin{equation}
r_1 = \frac{Ar}{2}
\end{equation}
\begin{equation}
r_2 = 2Ar
\end{equation}
where $A$ is additional multiplicational factor that was changed from 1 to 0.8 and 1.2. So we used parameter A to set the values of $r_1$ and $r_2$.

2.  We change the quantity of galaxies in a cluster - $n$. Initial value of $n$ is given by formula
\begin{equation}
n = \frac{z^5}{1000}
\end{equation}
where $z$ is number of filaments that are connected to a cluster. This value was multiplied by factor $B$ close to unity.

3. $I$ - number of field galaxies including isolated.

4. $F$ - number of galaxies in filament. This value is proportional to mass of filament.

5. $D$ -  radius of cluster. The distance of galaxy from cluter center was calculated as tangent of random number from 0 to $\pi /2$ multiplied to radius coefficient D. So the half of cluster galaxies lie at the distance no more then D from center. Our initial value for D was 1.4 degrees.

6. $N$ - total number of clusters.

Fig. 1 demonstrate random distribution of points, which coordinates depends on distance between this points and are connected with straight filaments. There are some tendency towards the accumulation and void formation as we can see from this picture of random points distribtuion. At first we simulated distribution according to initial conditions which reflect cellular structure of Universe. Results obtained by such way are demonstrated at Fig. 1. As we can see from this graph, result is quite rough and does not reflect real picture of situation. The second stage of work was the attempt to simulate large-scale structure of Universe with the help of random distribution of single galaxies. 

We used the distribution of galaxies modeled using Gaussian random fields depending on the mass of filaments (the number of galaxies present there) and boundary distances between them, with addition of isolated galaxies. The resulting distribution is presented on Fig. 2. 

Fig. 3 represent observable distribution of galaxies based on the SDSS sampling for radial velocity range from 4000 to 11000 km/s. Comparing it with Fig. 2 one can see unquestioning similarity,that can be estimated quantitatively, by calculating the angular two-point correlation function. We used Landy-Szalay estimator (4) for calculate correlation between SDSS and our models (Vargas-Megana, 2012).
\begin{equation}
\xi = \frac{DD-2DR+RR}{RR}
\end{equation}

DD is the number of pair of galaxies in certain angular range in the considered sample, RR is the same value for randimised sample and DR is the number of pairs of one real galaxy and one point from randomised sample. We approximate the correlation function by power law with two parameters: power index $\alpha $ and normalisation factor. In Table 1, there are demonstrated different values of power indicator and rate of normalization, that was obtained by changing several parameters in our model. If mass of filament, isolated galaxies quantity and radius of cluster are reduced by 40\%, we got power indicator similar to SDSS survey. Fig. 4 demonstrate how this reduced model looks like. Visualization effect is responsible for such unlikely appearance, Fig. 5 shows enlarged part of this picture with the following coordinate ranges: 14<x<15.5 and 45<y<55. X coordinate corresponds to right ascension in hours and Y corresponds to declination in degree for obvious comparison with SDSS galaxy distribution. It should be noted that the simulation of other observational samples, including SDSS galaxies at larger distances, should apply different parameters or even some algorithm details.\\[2mm]

{\bf 3. Results and conclusions}\\[1mm]

We had simulated the matter distribution that in result gives results close to observed. 
This two dimensional visualisation in its general features is similar to real structure of filaments, galaxy clusters and voids. 
We compared our model with SDSS catalog by angular two-point correlation function. 
Changing initial conditions led our distribution to quantitive similarity with SDSS observable data.\\[2mm]

\begin{table}
 \centering
 \caption{Approximation parameters.}\label{tab1}
 \vspace*{1ex}
 \begin{tabular}{lllll}
   \hline
 Sample & SDSS & M1 & M2 (-40\% ) & M3 (-45\% ) \\ \hline
 $\alpha $ & -1.4963 & -1.2747 & -1.4743 & -1.5228 \\
 Norm & 11.15 & 11.48 & 10.93 & 11.24 \\ 
 F & - & 5 & 3 & 2.75 \\
 I & - & 32400 & 19440 & 17820 \\
 D & - & 2 & 1.2 & 1.1 \\ \hline
  \hline 
 \end{tabular}
\end{table}

\begin{table}
 \centering
 \caption{Changed parameters.}\label{tab1}
 \vspace*{1ex}
 \begin{tabular}{ccc}
   \hline
Parameter change & $\alpha $ & Normalization \\ \hline
 $N$ -20\% & -1.2420  & 9.304 \\
 $N$ +20\% & -1.2532  & 9.559 \\ 
 $A$ -20\% & -1.2115  & 9.54 \\
 $A$ +20\% & -1.1892  & 8.835 \\ 
 $F$ -20\% & -1.2756  & 9.548 \\ 
 $F$ +20\% & -1.2274  & 9.371 \\ 
 $B$ -20\% & -1.275   & 9.763 \\ 
 $B$ +20\% & -1.1761  & 9.064 \\ 
 $D$ -20\% & -1.2964  & 9.722 \\ 
 $D$ +20\% & -1.2137  & 9.247 \\ 
 $I$ -20\% & -1.2847  & 9.723 \\
 $I$ +20\% & -1.244   & 9.366 \\ \hline
 \hline 
 \end{tabular}
\end{table}

%

%
{\bf References\\[2mm]}
Doroshkevich~A.\,G., Kotok~E.\,V., Poliudov~A.\,V. et. al. 1980, {\it MNRAS}, {\bf 192}, 321 \\
Kim~J.\,P., Park~C., Gott~J.\,R.., Dubinski~J. 2009, {\it ApJ}, {\bf 701}, 1547 \\
Tugay A. 2012. {\it Odessa Astron. Publ.,} {\bf 25}, 142 \\
Vargas-Megana M., et al. \ arXiv:1211.6211v2 \\
\vfill
%

%
%
\end{document}